\documentclass[epj]{svjour}
\usepackage{amsmath}
\usepackage{amssymb}
\usepackage{graphicx}
\usepackage{dcolumn}
\usepackage{bm}
\usepackage{dsfont}
\usepackage{color} 
\usepackage{hyperref}
\hypersetup{colorlinks,bookmarksopen,bookmarksnumbered,citecolor=blue, linkcolor=black,pdfstartview=FitH,urlcolor=blue}
\newcommand{\changes}[1]{#1}
\begin{document}
\title{Statistical interpretation of sterile neutrino oscillation searches at reactors}
\author{Pilar Coloma\inst{1}\thanks{\emph{Email:} pilar.coloma@ift.csic.es}%
  \and Patrick Huber\inst{2}\thanks{\emph{Email:} pahuber@vt.edu}%
  \and Thomas Schwetz\inst{3}\thanks{\emph{Email:} schwetz@kit.edu}%
}                     
%
%
\institute{Instituto de F\'isica Te\'orica UAM/CSIC, Calle Nicol\'as Cabrera 13-15, Universidad Aut\'onoma de Madrid, 28049 Madrid, Spain
  Instituto de F\'isica Corpuscular UV/CSIC, Calle Catedr\'atico Jos\'e Beltr\'an 2, Parque Cient\'ifico, E-46980 Paterna, Spain \and
Center for Neutrino Physics, Physics Department, Virginia Tech, Blacksburg, USA \and%
Institut f\"ur Astroteilchenphysik, Karlsruher Institut f\"ur Technologie (KIT), D-76021 Karlsruhe, Germany}
\date{Received: date / Revised version: date}
%
\abstract{
A considerable experimental effort is currently under way to test the persistent hints for oscillations due to an eV-scale sterile neutrino in the data of various reactor neutrino experiments. The assessment of the statistical significance of these hints is usually based on Wilks' theorem, whereby the assumption is made that the log-likelihood is $\chi^2$-distributed. However, it is well known that the preconditions for the validity of Wilks' theorem are not fulfilled for neutrino oscillation experiments. In this work we derive a simple asymptotic form of the actual distribution of the log-likelihood based on reinterpreting the problem as fitting white Gaussian noise. From this formalism we show that, even in the absence of a sterile neutrino, the expectation value for the maximum likelihood estimate of the mixing angle remains non-zero with attendant large values of the log-likelihood. Our analytical results are then confirmed by numerical simulations of a toy reactor experiment. Finally, we apply this framework to the data of the Neutrino-4 experiment and show that the null hypothesis of no-oscillation is rejected at the 2.6\,$\sigma$ level, compared to  3.2\,$\sigma$ obtained under the assumption that Wilks' theorem applies. 
%
} 
\maketitle
%
\section{Introduction}

Despite the strong evidence for the existence of physics beyond the
Standard Model (SM), searches for new particles at high-energy
colliders have been so far unsuccessful. A possible explanation is
that the new physics lies at relatively low scales, and that the dark
and the visible sectors communicate through a very weakly-interacting
particle. A particularly appealing scenario in this context is the
addition of right-handed neutrinos to the SM, which could also source
the SM neutrino masses. While right-handed neutrinos are singlets of
the SM gauge group (and therefore \emph{sterile} under the SM
interactions), they may lead to observable signatures through their
mixing with the SM neutrinos at a variety of experiments, depending on
their masses.  In particular, and motivated by the LSND
anomaly~\cite{Aguilar:2001ty}, considerable experimental effort was
directed towards the search for eV-scale right-handed neutrinos using
short-baseline oscillation experiments over the past two decades, see
Ref.~\cite{Boser:2019rta} for a recent review. In this article we
focus on reactor neutrino experiments, which have been playing a
crucial role in neutrino physics since the discovery of the neutrino
by Reines and Cowan~\cite{Cowan:1992xc}, and have been central to the
search for eV-scale sterile neutrinos.

Calculations of the anti-neutrino fluxes emitted from nuclear reactors
performed in 2011 \cite{Mueller:2011nm,Huber:2011wv} have lead to the
so-called ``reactor anti-neutrino anomaly''~\cite{Mention:2011rk},
providing a hint for the existence of sterile neutrinos due to an observed 
mis-match between predicted and observed rates. This hint remains
controversial up to today, see
Refs.~\cite{Berryman:2019hme,Giunti:2019qlt} for discussions of the
latest developments on this issue. Therefore, modern experiments
focus on the relative comparison of measured spectra at different
baselines \cite{An:2016luf,Choi:2020ttv,DoubleChooz:2019qbj,Ko:2016owz,Alekseev:2018efk,Andriamirado:2020erz,AlmazanMolina:2019qul,Serebrov:2018vdw}, which is more robust against
uncertainties in flux predictions. Recent global fits of reactor data
still find indications for sterile neutrino oscillations at the level
of $2-3\sigma$, even when using only spectral ratios
\cite{Dentler:2017tkw,Dentler:2018sju,Gariazzo:2018mwd,Berryman:2020agd}.
Those indications are based purely on spectral distortions which may
feature oscillatory patterns.

Note, however, that the necessary conditions to apply Wilks'
theorem~\cite{Wilks:1938dza} are typically not fulfilled for sterile
neutrino searches in oscillation experiments, which can lead to wrong
results when evaluating significance or confidence levels based on
$\chi^2$ values.\footnote{For similar studies in the context of
  three-flavor neutrino oscillations see for instance
  \cite{Schwetz:2006md,Blennow:2013oma,Blennow:2014sja,Elevant:2015ska,Esteban:2016qun}.}  Therefore, one should wonder
if the hints may be coming from a mis-interpretation of the data. This
has recently been pointed out in Refs.~\cite{Agostini:2019jup,Giunti:2020uhv,Almazan:2020drb}, where the authors have shown by
way of Monte Carlo simulations that that there are indeed corrections
and that the statistical significance of the hints is reduced. In this
paper we attempt to go one step further and, besides providing
analytical arguments that allow to understand the expected
distribution for the test statistics, we also study the dependence of
the observed corrections on relevant experimental parameters and
numerical details of the analysis.

The outline of the paper is as follows. In Sec.~\ref{sec:remarks} we start
with some general remarks, introduce a test statistic to evaluate the
significance of the presence of sterile neutrino oscillations and give
some qualitative arguments why it is likely that experiments find a
hint for sterile neutrinos even if there are none. In Sec.~\ref{sec:toy}
we consider an idealized disappearance experiment and derive the
expected distribution of the test statistics. We show that the above
statement is a consequence of the fact that in a Fourier composition
of white noise some frequency will appear with largest amplitude. This
will allow us to make predictions for the expected distribution of the
best fit points for $\sin^22\theta$. In Sec.~\ref{sec:simulation} we
perform numerical simulations of toy reactor experiments, and we study
in detail the distribution of the test statistic as well as the
location of the best-fit points. We also investigate the impact of
various parameters, such as restriction to the physical region, the
impact of systematics, or alternative $\chi^2$ definitions. In
Sec.~\ref{sec:conclusions} we consider as a case study the recent results
for the Neutrino-4 experiment
\cite{Serebrov:2020kmd,Serebrov:2018vdw}, which has reported a $\sim
3\sigma$ hint for sterile neutrino oscillations. We perform a
Monte-Carlo study of the Neutrino-4 data and show that the
significance for the presence of sterile neutrinos is somewhat lower
than expected under the assumption of a $\chi^2$ distribution of
the test statistics, and we present the confidence regions obtained by
explicit Neyman-Pearson construction based on the Feldman-Cousins prescription
\cite{Feldman:1997qc}. We conclude in Sec.~\ref{sec:conclusions}.

\section{General remarks}
\label{sec:remarks}

Reactor neutrino experiments look for the disappearance of
electron anti-neutrinos. In this work we assume that a single 
sterile neutrino is relevant for the phenomenology and we
consider only experimental setups where the baseline is short enough such that
oscillations due to the standard three-flavor mass-squared
differences can be safely neglected. In this limit, sterile neutrino
oscillations are described by an effective two-flavor survival
probability:
\begin{equation}\label{eq:Posc}
    P^{\rm osc} = 1 - \frac{1}{2}\sin^22\theta \left(1 - \cos\frac{\Delta m^2 L}{2 E} \right)  \,,
\end{equation}
where $E$ is the neutrino energy, $L$ is the baseline, $\theta$ is an
effective neutrino mixing angle, and $\Delta m^2$ stands for the mass-squared splitting between the eV-scale mass state and the light SM neutrinos.

In order to analyze the results of a given experiment a least-squares
function of binned spectral data
is considered, $\chi^2(\sin^22\theta,\Delta m^2)$. 
A common test statistic $T$ for evaluating the
hypothesis of the presence of sterile neutrino oscillations is the
$\Delta\chi^2$ (or, equivalently, the likelihood ratio) between the
best-fit point and the no-oscillation case:
%
\begin{align}
  T &= \chi^2(\text{no osc}) - \chi^2(\text{best fit}) \nonumber\\
    &= \chi^2(0,0) - \chi^2(\widehat{\sin^22\theta},\widehat{\Delta m^2}) \,,
    \label{eq:defT}
\end{align}
where $\widehat{\sin^22\theta}$ and $\widehat{\Delta m^2}$ indicate
the parameter values at the $\chi^2$ minimum. If Wilks'
theorem~\cite{Wilks:1938dza} applies, $T$ should be distributed as a
$\chi^2$-distribution for 2 degrees of freedom (DOF), corresponding to
the two minimized parameters.

Indeed, for the problem at hand, there are several reasons to suspect
that the necessary conditions for Wilks' theorem to apply are not
fulfilled: First, there is a physical boundary for the mixing angle,
$\sin^22\theta\ge 0$. Second, the parameter $\Delta m^2$ becomes
undefined for $\sin^22\theta\to 0$ and $\sin^22\theta$ becomes
unphysical for $\Delta m^2\to 0$. Third, the cosine dependence on
$\Delta m^2$ of the oscillation probability in eq.~\ref{eq:Posc} leads to a
strong non-linear behavior. Therefore, significant deviations of the
distribution of $T$ from a $\chi^2$-distribution are expected \emph{a
  priori}, see also \cite{Feldman:1997qc,Agostini:2019jup}. \changes{A recent review discussing the applicability of Wilk's theorem can be found in Ref.~\cite{Algeri:2020pql}.}

As we will show in Sec.~\ref{sec:toy}, for an idealized situation the
distribution of $\sqrt{T}$ is the one of the maximum of $N$ standard normal random
variables, where $N$ corresponds to an effective number of bins. We
will give physical arguments, as to why for this type of experiments a
non-vanishing value for $\sin^22\theta$ at the best-fit is likely,
with a relatively large value of $T$. In fact, its typical value is
set by the size of the relative statistical uncertainty of the
sample. In Sec.~\ref{sec:simulation} we will compute the distribution of
$T$ for more realistic configurations and will always confirm rather
large deviations from a $\chi^2$-distribution. This suggests that
reliable statements about significance and confidence levels require
explicit Monte-Carlo simulations, in agreement with previous
results~\cite{Agostini:2019jup,Giunti:2020uhv}. We will demonstrate
this explicitly using the recent results from Neutrino-4 in
Sec.~\ref{sec:neutrino4}.

\section{Expected distribution of the test statistics}
\label{sec:toy}

In this section we derive the expected test statistic $T$ for a toy model. After making a series of assumptions that allow us to write the randomly fluctuated events in each bin as a discrete Fourier transform, we will proceed and minimize the $\chi^2$ function analytically. This will provide us with an expression in terms of the Fourier coefficients of the expansion, which we can then substitute into eq.~\ref{eq:defT} to get an analytical expression for the expected test statistic. 

\subsection{Derivation of the test statistic for a toy model}
\label{sec:assumptions}

Let us consider a toy model for an oscillation disappearance experiment. We consider $N$ bins in $L/E$ and write the predicted event number in each bin as
\begin{equation}
    p_i \approx p_i^0 P^{\rm osc}_i \,,
\end{equation}
where $p_i^0$ is the predicted number of events in case of no oscillations,  
$P^{\rm osc}_i$ is given in eq.~\ref{eq:Posc}, and the index $i$ labels the bins in $L/E$.
Let us now adopt the following assumptions:
\begin{enumerate}
    \item[$(a)$] 
    We assume that there is \textbf{no sterile neutrino} in Nature, i.e., the observed data in bin $i$ is given by the no-oscillation prediction plus a statistical fluctuation with variance $\sigma_i^2$:
\begin{align}
    &d_i = p_i^0 + \delta d_i  \quad\text{with} \nonumber\\ 
    &\langle \delta d_i \rangle = 0\,,\quad 
  \langle \delta d_i \delta d_j \rangle = \sigma_i^2 \delta_{ij} \,.
  \label{eq:data}
\end{align}
For Poisson statistics we would have 
$\sigma_i = \sqrt{p_i^0}$. For simplicity we assume that $\delta d_i$ are Gaussian.

\item[$(b)$] We assume that \textbf{only shape information in $L/E$} is used, but not on the absolute normalization. This applies to experiments where energy spectra are fitted leaving the overall normalization free, but also to setups or combinations of experiments where relative spectra at different baselines are considered.
\end{enumerate}

The second assumption is implemented in a $\chi^2$ by introducing a free pull parameter $\xi$, such that
\begin{align}\label{eq:chi2_gauss}
     \chi^2 = \sum_{i=1}^N \left[\frac{d_i - (1+\xi)p_i}{\sigma_i}\right]^2\,,
\end{align}
where minimization with respect to $\xi$ is understood. Working to linear order in $\xi$ and $\sin^22\theta$, and assuming that terms involving a sum over $\delta d_i$ or $\cos (\Delta m^2L/2E_i)$ average to zero, one finds that $\xi \approx (1/2) \sin^22\theta$ minimizes the $\chi^2$. Using this together with assumption $(a)$ above we find
\begin{equation}\label{eq:chi2_1}
    \chi^2 = \sum_{i=1}^N \left[n_i - \frac{p_i^0}{2\sigma_i} \sin^22\theta  \cos\frac{\Delta m^2 L}{2 E_i} \right]^2  \,,
\end{equation}
where $n_i \equiv \delta d_i/\sigma_i$ are independent standard normal random variables with
$\langle n_i\rangle = 0$, $\langle n_i n_j\rangle = \delta_{ij}$, see eq.~\ref{eq:data}. 

Let us now adopt the additional simplifying assumptions to build a mathematical toy model:
\begin{enumerate}
    \item[$(c)$] We assume that the relative statistical \changes{uncertainty} $\sigma_i/p_i^0$ has the same value for each bin and define the new parameter 
    \begin{equation}\label{eq:defa}
        s \equiv \frac{p_i^0}{2\sigma_i} \sin^22\theta \,.
    \end{equation}
\end{enumerate}
Although this is not strictly the case for a reactor experiment, in Sec.~\ref{sec:simulation} we will see that it works relatively well for the experimental setups under consideration in this work. Furthermore we assume that bins have equal width in $L/E$ and define
    \begin{equation}\label{eq:bins_discrete}
        \frac{\Delta m^2}{2} \left(\frac{L}{E}\right)_j = 
        \frac{2\pi}{N} \kappa j \equiv \varphi_{\kappa j} \,.
    \end{equation}
Hence, $j$ labels bins in $L/E$ while the index $\kappa$ labels discrete frequencies proportional to $\Delta m^2$. With this idealization, eq.~\ref{eq:chi2_1} becomes
\begin{equation}\label{eq:chi2_2}
    \chi^2(s,\kappa) = \sum_{i=1}^N \left[n_i - s \cos\varphi_{\kappa i} \right]^2  \,.
\end{equation}
We see that in this limit the sterile neutrino search is equivalent to fitting  Gaussian white-noise with a cosine function with the amplitude $s$ and the frequency $\kappa$ as free parameters. This form suggests to consider the discrete Fourier transform of the $N$ random variables $n_i$:
\begin{equation}\label{eq:FTn}
n_i = \sum_{\kappa = 1}^N \left( a_\kappa \cos\varphi_{\kappa i} + b_\kappa\sin\varphi_{ \kappa i} \right)
\end{equation}
with $a_\kappa, b_\kappa \in \mathds{R}$. Focusing on the cosine term, the coefficients $a_\kappa$ can be computed as
\begin{equation}
    a_\kappa = \frac{2}{N} \sum_{i=1}^N n_i \cos\varphi_{\kappa i} \,.
\end{equation}
Since $n_i$ are independent standard Gaussian variables, it is clear that $a_\kappa$ are random Gaussian variables as well, with
\begin{equation}\label{eq:ak}
  \langle a_\kappa \rangle = 0 \,,\quad \langle a_\kappa a_\lambda \rangle = \frac{2}{N} \delta_{\kappa \lambda} \,,   
\end{equation}
where we have assumed that sums over  $\cos\varphi_{\kappa i}$ $(\cos^2\varphi_{\kappa i})$ average to 0 ($N/2$).

Let us now look for the best fit point ($\hat s, \hat \kappa$). We start by minimizing the $\chi^2$ in eq.~\ref{eq:chi2_2} with respect to $s$, for fixed $\kappa$. This gives
\begin{equation}
    \hat s(\kappa) = \frac{\sum_{i=1}^N n_i \cos\varphi_{\kappa i}}{\sum_{i=1}^N \cos^2\varphi_{\kappa i}} = a_\kappa \,,
\label{eq:ahatk}
\end{equation}
where in the last step we have inserted $n_i$ from eq.~\ref{eq:FTn}, using the fact that all terms average to zero except the one containing $\cos^2\varphi_{\kappa i}$. This implies that, for fixed $\kappa$, $\hat s(\kappa)$ follows a Gaussian distribution with its mean and variance as given by eq.~\ref{eq:ak}. Also, we see that, for fixed $\kappa$, the $\chi^2$ is minimized by choosing $s$ as the Fourier coefficient corresponding to the frequency $\kappa$. 
Inserting the Fourier transform from eq.~\ref{eq:FTn} as well as the solution from eq.~\ref{eq:ahatk} into the $\chi^2$ in eq.~\ref{eq:chi2_2}, we find
\begin{align}
    \chi^2(\hat s(\kappa), \kappa) &= \sum_{i=1}^N \left[
      \sum_{\lambda \neq \kappa} a_\lambda \cos\varphi_{\lambda i} + \sum_\lambda b_\lambda \sin\varphi_{\lambda i} \right]^2 \nonumber\\
    &= \frac{N}{2} \sum_{\lambda \neq \kappa} a_\lambda^2 + C \,.
    \label{eq:chi2_3}
\end{align}
where in the second step we have expanded the square and used the fact that only terms of the form $\cos^2\varphi_{\lambda i}$ or $\sin^2\varphi_{\lambda i}$ survive, while all mixed terms average to zero. Here, the constant $C$ contains the $b_\lambda$ terms and is independent of $\kappa$.


Next, we minimize with respect to $\kappa$. Since eq.~\ref{eq:chi2_3} is a sum of positive terms, the $\chi^2$ would be minimal for $\kappa = \hat \kappa$, such that $a_{\hat \kappa}$ is the Fourier coefficient with the largest absolute value. However, considering the definition of $s$ in eq.~\ref{eq:defa}, we see that the physical requirement $\sin^22\theta \ge 0$ implies $s\ge 0$. Therefore, if the minimization is restricted to the physically allowed region we obtain
\begin{equation}\label{eq:ahat}
    \hat s \equiv \hat s (\hat \kappa) = a_{\hat \kappa} = \max[ 0, \max_\kappa a_\kappa] \,.
\end{equation}
For $N$ Gaussian variables with $\langle a_\kappa \rangle =0$, the probability that all $a_\kappa$ are negative is $(1/2)^N$. Hence, for sufficiently large $N$ it is very likely to obtain at least one positive $a_\kappa$, such that eq.~\ref{eq:ahat} leads to a positive best-fit point for the parameter $s$ (and therefore for $\sin^22\theta$). For simplicity we neglect hereafter the unlikely case that none of the $a_\kappa$ is positive.

Finally, let us consider the test statistic $T = \chi^2(0,0) - \chi^2(\hat s, \hat \kappa)$ defined in eq.~\ref{eq:defT}. Using eqs.~\ref{eq:chi2_2} and \ref{eq:FTn} we find the $\chi^2$ for the SM point as
\begin{equation}
    \chi^2(0,0) = \sum_i (n_i)^2 = \frac{N}{2} \sum_\lambda a_\lambda^2 + C \,,
\end{equation}
where $C$ is the same constant as in eq.~\ref{eq:chi2_3}. Evaluating now the minimum of the $\chi^2$, $\chi^2(\hat s, \kappa)$ as given in eq.~\ref{eq:chi2_3} at $k = \hat \kappa$, we finally obtain
\begin{equation}\label{eq:Ttoy}
    T = \frac{N}{2} a_{\hat \kappa}^2 = \left[ \max_\kappa \tilde a_\kappa \right]^2 \,,
\end{equation}
where $\tilde a_\kappa \equiv \sqrt{N/2} a_\kappa$ are standard normal random variables, 
$\langle \tilde a_\kappa\rangle = 0$, $\langle \tilde a_\kappa\tilde a_\lambda \rangle = \delta_{\kappa\lambda}$ (see eq.~\ref{eq:ak}).  

\subsection{Discussion}

Equations~\ref{eq:ahat} and \ref{eq:Ttoy} are the main results of this section. The latter shows that the square-root of the test statistic $T$ has the distribution of the maximum of $N$ standard normal variables. It is proportional to the best fit amplitude $\hat s$, and
hence, up to a normalization factor, the best-fit point in eq.~\ref{eq:ahat} follows the same distribution. Distributions of this type are considered in the field of ``extreme value statistics'', see e.g., \cite{coles,pfeifer}.

For the case of Gaussian variables of interest here, there exists a limiting distribution for $N\to\infty$. It is based on the so-called \emph{Gumbel} distribution $e^{-e^{-z}}$. Let $x = \max_i \tilde a_i$, where $\tilde a_i$ are $N$ standard normal variables. For finite $N$ the cumulative probability distribution (CDF) $F(x)$  can be approximated by \cite{pfeifer}: 
\begin{equation}\label{eq:gumbel}
    F(x) = \exp\left\{-\exp\left[-A_N(x - B_N)\right]\right\}  \,,
\end{equation}
with
\begin{equation}
    A_N = \sqrt{2\log N} \,,\quad
    B_N = A_N - \frac{\log\log N + \log 4\pi}{2 A_N} \,.
\end{equation}

\begin{figure*}
\begin{center}
\includegraphics[width=0.9\textwidth]{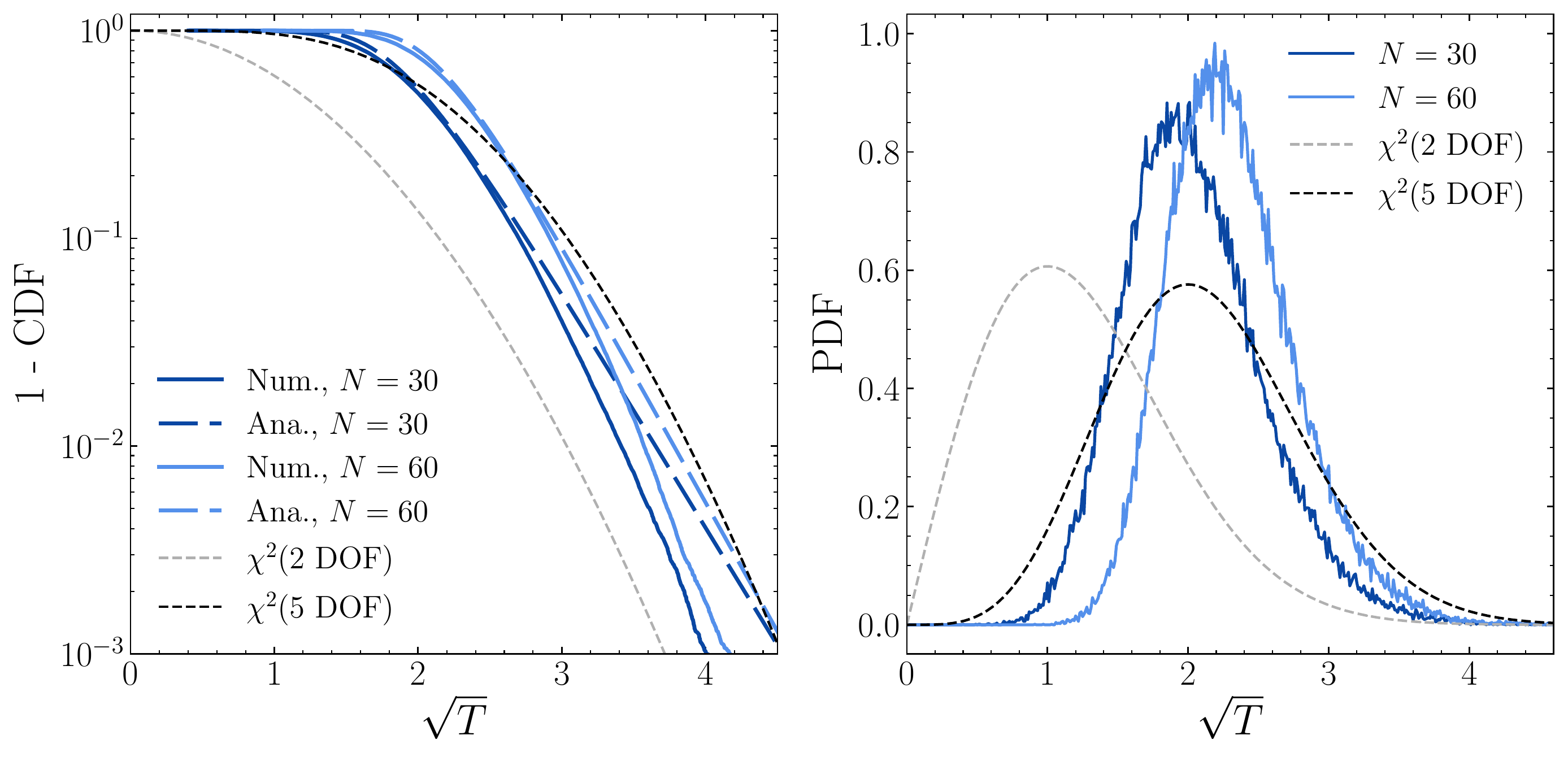}
\caption{Expected distribution for the square-root of the test statistic $\sqrt{T} = \max_\kappa \tilde a_\kappa$, where $\tilde a_\kappa$ are $N$ standard normal random variables. The left panel shows 1--CDF, while the PDF is shown in the right panel. Dark (light) blue curves correspond to $N=30\,(60)$. Solid curves are obtained by numerical simulations, whereas long-dashed curves correspond to the approximation in eq.~\ref{eq:gumbel}. For comparison, the short-dashed gray and black curves show the distributions obtained if $T$ would follow a $\chi^2$ distribution for 2 and 5 DOF, respectively.}
\label{fig:analytical} 
\end{center}
\end{figure*}

In Fig.~\ref{fig:analytical} (left) we show 1--CDF for the maximum of $N$ standard normal variables obtained by numerical calculations (solid) compared to the approximate formula in eq.~\ref{eq:gumbel} (long-dashed) for $N=30$ and 60. We see that they agree reasonably well for $1-\text{CDF} \gtrsim 0.1$, but start to deviate for smaller values. Indeed, the convergence to the Gumbel distribution goes only as $1/\log N$ \cite{pfeifer}. Therefore, since the distribution can be easily calculated numerically, we will for the rest of the paper stick to the numerical method and denote this distribution by ``Max.~Gauss'' in the following.

The important property of this distribution is, that small values of $T$ are rather unlikely. In 
Fig.~\ref{fig:analytical} we compare 1--CDF as well as the probability density function (PDF) for $\sqrt{T}$ to the one \changes{for the square-root} of a $\chi^2$ distribution.\footnote{\changes{Note that the Jacobian of the variable transformation has to be taken into account when transforming the PDF for $\chi^2$ into the PDF for $\sqrt{\chi^2}$.}} Indeed, if Wilks' theorem was applicable, $T$ should be distributed as $\chi^2$ with 2 DOF. Obviously, the conditions for Wilks' theorem to hold are badly violated in this case, for the reasons mentioned in Sec.~\ref{sec:remarks}. The peak at $\sqrt{T} \sim 2$ and the small probability to obtain $\sqrt{T} \lesssim 1$ indicates that, even if there is no sterile neutrino present in Nature, it is very likely to obtain a best-fit point with finite $\sin^22\theta$ as well as relatively large value of $T$. This would lead to claiming a signal at relevant statistical significance, if evaluated with a $\chi^2$-distribution. The physical reason for this behavior can be understood from eq.~\ref{eq:chi2_2}: in a white noise spectrum it is very likely to find some frequency with sizable amplitude that is able to fit the data. 

The expectation value for a random variable $z$ with the CDF $F(z) = e^{-e^{-z}}$ is given by \cite{pfeifer} $\langle z\rangle = \gamma$, where $\gamma=0.57721\ldots$ is the Euler-Mascheroni constant. From eqs.~\ref{eq:Ttoy}, \ref{eq:gumbel} follows then 
\begin{equation}\label{eq:Texpect}
    \left\langle\sqrt{T}\right\rangle = B_N + \frac{\gamma}{A_N} \approx 2\ldots 2.4 \,,
\end{equation}
where the numbers hold for $N \approx 30\ldots 60$. These values agree to a good accuracy with the mean values obtained numerically, and depend only weakly (logarithmically) on $N$. We can use these results to estimate the expectation value for $\sin^22\theta$. Let $\mathcal{N}$ be the total number of observed events. According to assumption ($c$) above we have $p_i^0 \approx \mathcal{N}/N$ and 
$\sigma_i = \sqrt{p_i^0}$. Then eq.~\ref{eq:defa} leads to
\begin{equation}\label{eq:Sq_expect}
     \sin^22\theta= 2 \sqrt{\frac{2}{\mathcal{N}}} \sqrt{T} 
     \quad\text{and}\quad
     \langle\sin^22\theta\rangle\approx \frac{6.2}{\sqrt{\mathcal{N}}}\,,
\end{equation}
where in the second relation we have used the numerical values from eq.~\ref{eq:Texpect}.
We see that up to a numerical factor, the expected best fit value for $\sin^22\theta$ is set by the relative statistical \changes{uncertainty} of the event sample. We will find this behavior in the simulations discussed in the following sections. From Fig.~\ref{fig:analytical} we see that there is a lower bound of $\sqrt{T} \gtrsim 1.5$ at 99\%~CL for $N=30\ldots 60$, which translates into a lower bound on $\sin^22\theta$ according to eq.~\ref{eq:Sq_expect}. 

To conclude this section, we remark that the idealized situation considered here is certainly an over-simplification, and especially assumption $(c)$ will not be satisfied in a realistic oscillation experiment. Nevertheless, these considerations capture the most relevant features and the results obtained here allow an intuitive understanding of the numerical results we are going to present below. In particular, the preference for the presence of sterile neutrino oscillations even in case of no true signal is predicted from those arguments, and allows a qualitative (in some cases even quantitative) understanding of the more realistic simulations discussed in the remainder of this paper. 

\section{Numerical simulations for a toy experiment}
\label{sec:simulation}

\subsection{Description of the simulation}
\label{sec:sim-details}

In order to verify the validity of the analytical approach presented in the previous section, we now proceed to perform a numerical simulation for a toy experiment. For this purpose, we choose a reactor disappearance experiment which aims to set a constraint on the $\sin^22\theta - \Delta m^2$ parameter space from the observation of $\bar\nu_e \to \bar\nu_e$ oscillations. We consider generic shapes of the anti-neutrino flux and inverse beta-decay detection cross section. The distance between the reactor core and the detector is set to $L=10$~m. In order to account for the finite size of the reactor core, the probability is averaged over a window $\Delta L = \pm 1$~m:
\begin{equation}
\label{eq:Pee_avg_DeltaL}
    \langle P^{\rm osc}(\theta, \Delta m^2) \rangle = 
    1 - \sin^22\theta \,
    \frac{  \int_{L-\Delta L}^{L + \Delta L} dL' 
    \sin^2(\frac{\Delta m^2 L'}{4E} )/L'^2 }{ \int_{L - \Delta L}^{L + \Delta L} dL' \frac{1}{L'^2}} \,.
\end{equation}
This ensures that fast oscillations are averaged-out at the detector.
Unless otherwise stated, the exposure is set such that the total number of events is $1.5\times 10^4$. A binned $\chi^2$ analysis is performed, using 43 bins in energy of equal size distributed between 2 and 8~MeV, and a Gaussian energy resolution of the form $ \sigma(E) = 0.03\sqrt{E/{\rm MeV}}$ is applied to the event distributions. 

The experimental details outlined above have been chosen to lie in the same ballpark as for some of the running short baseline reactor disappearance experiments \cite{Ko:2016owz,Alekseev:2018efk,Andriamirado:2020erz,AlmazanMolina:2019qul,Serebrov:2018vdw}. However, we have explicitly checked that changing any of these parameters does not qualitatively affect our results. Finally, we have assumed negligible backgrounds in our analysis for simplicity, in order to ease the interpretation of our results. Again in this case, we have checked that the inclusion of a sizable background component does not alter qualitatively our conclusions.

The results presented in this section have been obtained by simulating a large sample of pseudo-experiments, applying random statistical fluctuations to the expected event rates for the reactor experiment setup outlined above. Since here we are mostly interested in evaluating the significance of a potential positive signal, throughout this section we will generate random data under the null-hypothesis, that there is no sterile neutrino in Nature, {\it i.e.}, for no oscillations. These are generated on a bin-per-bin basis, sampling a normal distribution with its mean set to the expected number of events in a given bin for the SM hypothesis $p_i^0 \equiv p_i(\theta = 0)$, and its width set to the associated statistical uncertainty $\sqrt{p_i^0}$.
For the large number of events considered here the Gaussian approximation to the Poisson distribution is well justified.
Unless otherwise stated, the number of pseudo-experiments simulated is set to 20,000 for each of the cases studied in this section.

The sample of pseudo-experiments will then be used to determine the distribution of our test statistics $T$ defined in eq.~\ref{eq:defT}. In order to do so, for each pseudo-experiment a Poisson $\chi^2$ function is built. For a set of parameters $(\theta, \Delta m^2)$, it reads \cite{Baker:1983tu}: 
\begin{align}
    \chi^2_{\rm stat, Poisson}&(\theta, \Delta m^2) =  
    2 \sum_i \Big[ (1 + \xi)\, p_i (\theta, \Delta m^2)  \nonumber\\ 
&- d_i - d_i \log \frac{(1 + \xi)\, p_i(\theta, \Delta m^2)  }{d_i}\Big]\,,
    \label{eq:chi2-stat-poisson}
\end{align}
where $p_i$ is the expected number of events in the $i$-th bin for $\theta $ and $\Delta m^2$ (in the absence of statistical fluctuations), while
$d_i$ is the ``observed'' number of events, i.e., the pseudo-data generated as described above. Here, $\xi$ is a nuisance parameter, introduced in order to account for the systematic uncertainty in the prediction of the expected event rates. Once eq.~\ref{eq:chi2-stat-poisson} has been computed, a pull-term is added and the result is minimized over the nuisance parameter $\xi$:
\begin{equation}
    \label{eq:chi2_total}
    \chi^2 (\theta, \Delta m^2) = \min_\xi \left[\chi^2_{\rm stat, Poisson}(\theta, \Delta m^2)   + 
    \left(\frac{\xi - \bar\xi}{\sigma_{\rm sys}}\right)^2 \right]
\end{equation}
where $\sigma_{\rm sys}$ stands for the prior uncertainty on the signal normalization. Here, $\bar\xi$ is a parameter introduced to account for the fact that the normalization of the signal is typically obtained from previous experimental data, which is also subject to statistical fluctuations. In order to account for the associated uncertainty, for each pseudo-experiment the value of $\bar\xi$ is drawn from a normal distribution centered at zero and with a width equal to $\sigma_{\rm sys}$, as for instance in Ref.~\cite{Blennow:2014sja}. 
For each pseudo-data realization we minimize the $\chi^2$ in  eq.~\ref{eq:chi2_total} with respect to $\sin^22\theta$ and $\Delta m^2$ and calculate a value for the test statistic $T = \chi^2(0,0) -\chi^2_{\rm min}$. From the ensemble of all simulated realizations we obtain then the expected distribution of $T$ under the null-hypothesis of no oscillations.

Eq.~\ref{eq:chi2-stat-poisson} will be our default $\chi^2$ definition. \changes{But we have also studied the case where the Poisson $\chi^2$ in eq.~\ref{eq:chi2-stat-poisson} is replaced by other commonly used $\chi^2$ definitions. The classical definition going back to Pearson \cite{pearson} is
\begin{equation}
    \label{eq:chi2-stat-pearson}
    \chi^2_{\rm stat, Pearson}(\theta, \Delta m^2) = \sum_i
    \left[ \frac{(1 + \xi)\, p_i(\theta, \Delta m^2) - d_i}{\sqrt{p_i(\theta, \Delta m^2)}}\right]^2 \, .
\end{equation}
In order to avoid the parameter dependence in the denominator, the variance is often estimated by the data itself. We denote this version in the following as Gauss $\chi^2$:
\begin{equation}
    \label{eq:chi2-stat-gauss}
    \chi^2_{\rm stat, Gauss}(\theta, \Delta m^2) = \sum_i
    \left[ \frac{(1 + \xi)\, p_i(\theta, \Delta m^2) - d_i}{\sqrt{d_i}}\right]^2 \,.
\end{equation}
As we will show below, the Gaussian $\chi^2$ can lead to different results for the distribution of $T$, while the Pearson definition leads to the same result as the Poisson case, eq.~\ref{eq:chi2-stat-poisson}, for sufficiently large event numbers per bin (as it is the case for the situations considered here).}

For the results presented in the following we assume a single baseline setup. However, the arguments presented in Sec.~\ref{sec:toy} apply also to multi-baseline configurations, for instance when ratios of spectra at different baselines are considered, or in case of segmented detectors with additional $L$ information. The derivation in Sec.~\ref{sec:toy} relies only on general binning in $(L/E)$ including also bins in $L$. We have verified explicitly that the simulation of a setup combining energy spectra at two different baselines leads to very similar results as the single-baseline configuration. This is also confirmed in Sec.~\ref{sec:neutrino4}, when we consider the Neutrino-4 experiment. 

\subsection{Results}

\begin{figure*}[t]
\begin{center}
\includegraphics[scale=0.5]{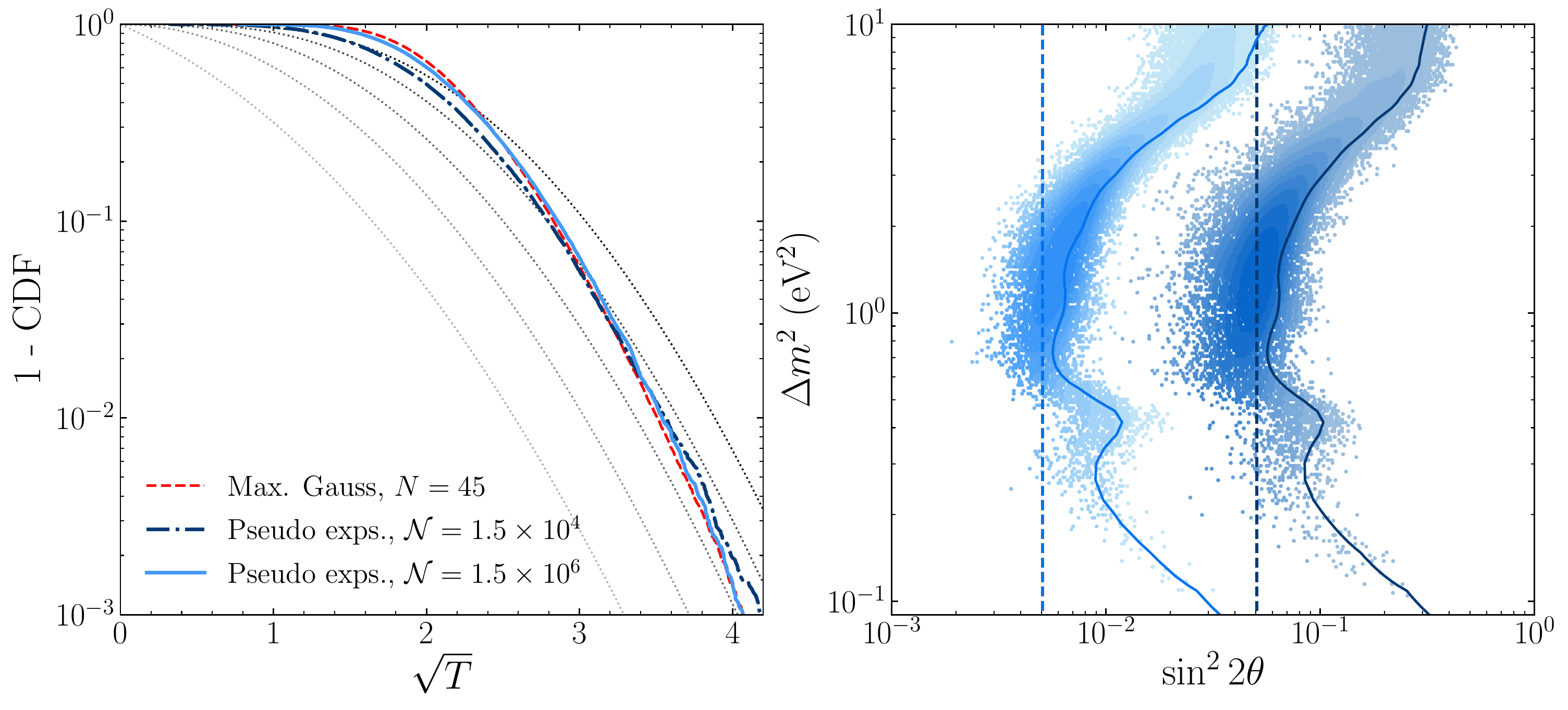}
\caption{\label{fig:bestfit} Left panel: Distribution of the test
  statistics obtained from numerical simulations for the toy reactor
  experiment described in the text. For comparison, the red-dashed curves shows the max.~Gauss
  distribution for $N=45$. Right panel: location of the
  best-fit points in the $\sin^22\theta - \Delta m^2$ plane, after
  minimization over nuisance parameters. In both panels, darker
  (lighter) blue lines/points correspond to the results obtained for
  $\mathcal{N} = 1.5\times 10^4 \, (1.5\times 10^6)$ events, using a
  sample of 20,000 pseudo-experiments simulated under the no-oscillation
  hypothesis. \changes{In the right panel, the regions with a higher density of best-fit points are indicated by the darker shades.} The dotted gray lines in the left panel show the
  $\chi^2$-distributions as the number of degrees of freedom is
  increased from 1 (lightest gray line to the left) to 5 (darkest gray
  line to the right). In the right panel, the vertical lines 
  indicate the predicted value of $\langle \sin^22\theta \rangle$ from
  eq.~\ref{eq:Sq_expect}. The solid curves show the expected 
  sensitivity at 95\% CL assuming that Wilks' theorem holds. 
  These results have been obtained using a
  Poisson $\chi^2$, with no background, for 10\% signal systematics,
  and restricting $0 < \sin^22\theta < 1$ in the fit. }
\end{center}
\end{figure*}

Fig.~\ref{fig:bestfit} presents the results of two simulations with different exposures: our default setup, with $\mathcal{N} =  1.5\times 10^4$ total number of events (dark blue), and the same setup with 100 times more events, $\mathcal{N} = 1.5\times 10^6$ (light blue). The left panel shows the distribution of the test statistic $T$. We observe a clear deviation from the $\chi^2$ distribution, and a good agreement with the max.~Gauss distribution derived in Sec.~\ref{sec:toy}. 

The agreement is excellent for the high statistics case and, in particular, we obtain the best match 
when the number of bins for the max.~Gauss distribution is set at $N=45$, to be compared with the 43 spectral bins used in the simulation. The reason for this (small) difference is that for a more realistic spectrum, some of the assumptions from Sec.~\ref{sec:toy} are only approximately fulfilled. In particular, assumption~($c$) (defined in Sec.~\ref{sec:assumptions}) requires that relative statistical \changes{uncertainties} are equal in all bins, which is obviously not true for a peaked spectrum as in reactor experiments. Therefore, $N=45$ should be considered as the \emph{effective} number of random standard normal variables, which leads to the best representation of the $T$ distribution from simulation.

\begin{table*}[ht!]
    \caption{Comparison of the confidence level (CL), $p$-value, and corresponding number of standard deviations ($\sigma$), for several values of $T$, obtained for a $\chi^2$ distribution with 2~DOF and for the max.~Gauss distribution for $N=45$. }
    \label{tab:Tvalues}
\renewcommand{\arraystretch}{1.3}
\setlength{\tabcolsep}{10pt}
    \centering
    \begin{tabular}{c@{\,\,}|lc|cc|lc}
    \hline\hline
     & \multicolumn{2}{c|}{CL [\%]}
     & \multicolumn{2}{c|}{$p$-value [\%]}
     & \multicolumn{2}{c}{\quad Number of $\sigma$} \\
    $T$ & $\chi^2(2)$ & max.~G. & $\chi^2(2)$ & max.~G. & $\chi^2(2)$ & max.~G. \\
    \hline
4.61 & 90.00  & 48.55 & 10.0 & 51.4 & 1.64 & 0.65\\
6.18 & 95.45  & 74.73 & 4.55 & 25.3 & 2.00 & 1.14\\
9.21 & 99.00  & 94.72 & 1.00 & 5.27 & 2.58 & 1.94\\
9.49 & 99.13  & 95.45 & 0.87 & 4.55 & 2.62 & 2.00\\
11.83 & 99.73 & 98.69 & 0.27 & 1.31 & 3.00 & 2.48\\
14.78 & 99.938& 99.73 & 0.062& 0.27 & 3.42 & 3.00\\
 \hline\hline
    \end{tabular}
\end{table*}
In Tab.~\ref{tab:Tvalues} we show, for various values of the test statistic $T$, the significance which would be obtained by assuming a $\chi^2$ distribution for 2~DOF (as we would expect if Wilks' theorem held) compared to the correct result following from the max.~Gauss distribution for $N=45$. For example, if a value of $T=11.83$ is observed, we would exclude the SM at $3\sigma$ ($p$-value 0.27\%) under the assumption of Wilks' theorem, while the correct significance would be only $2.48\sigma$ ($p$-value 1.3\%). As a rule of thumb, we can see from the table that $p$-values are under-estimated by about a factor 5, and the number of $\sigma$ gets reduced by roughly $0.5\sigma$ (except for low CL, where the difference is close to 1$\sigma$). 

The right panel in Fig.~\ref{fig:bestfit} shows the distribution of the best-fit points obtained in the simulations. Although the pseudo-data has been generated under the no-oscillation
hypothesis, we observe a clear preference for a non-vanishing value of $\sin^22\theta$. 
Note that actually \emph{none} of the best fit points is located near the ``true value'' ($\sin^22\theta=0$). Obviously $\widehat{\sin^22\theta}$ and $\widehat{\Delta m^2}$ are biased estimators in this case. 
Comparing the light and dark blue \changes{bands} we confirm the
scaling of the value of the mixing angle at the best fit with the relative statistical \changes{uncertainty}  
$1/\sqrt{\mathcal{N}}$, as expected from the discussion in
Sec.~\ref{sec:toy}. Indeed, in the region $0.5\,{\rm eV}^2 \lesssim \Delta
m^2 \lesssim 3\,{\rm eV}^2$ the mean value of the $\sin^22\theta$ best-fit points agrees rather well with the prediction from
eq.~\ref{eq:Sq_expect}, as indicated by the vertical dashed lines.\footnote{
We have verified that the range of values of $\Delta m^2$ where this is satisfied 
scales with the baseline as expected from $\Delta m^2 L = const$.} \changes{Interestingly, this region also contains the highest concentration of best-fit points (indicated by the darker shading in each case), while it is more difficult to obtain a result favoring larger/smaller values of the mass splitting. }
However, outside this region of $\Delta m^2$ the best-fit points lie at 
larger values for the mixing angle. The reason is that for extreme values of $\Delta m^2$
the idealizations assumed in Sec.~\ref{sec:assumptions} do not apply. For example, the feature
around $\Delta m^2 \sim 0.4$~eV$^2$ corresponds to $\Delta m^2 L/ (2E)
\simeq \pi$ at $E\simeq 3$~MeV and, as a result, the first minimum of the
survival probability is located at the peak of the event
spectrum. This corresponds roughly to the case where half an
oscillation period fits into the effective energy range, and therefore
corresponds to the minimal frequency which can be sampled by the
data. In contrast, for high mass-squared differences the frequency
becomes much higher than the bin width can capture and therefore
corresponds to over-sampling of the data.\footnote{\changes{Let us note that the optimal bin width should be determined by the energy resolution of the detector.}}
Hence, in both cases we are
leaving the domain of the discrete parameterization of $\Delta m^2$ in
terms of the index $\kappa=1, \ldots, N$ adopted in Sec.~\ref{sec:assumptions}, see
eq.~\ref{eq:bins_discrete}, which leads to the observed deviations with respect to the
estimate in eq.~\ref{eq:Sq_expect}.

For comparison, we also show in the right panel of Fig.~\ref{fig:bestfit} the expected sensitivity at 95\%~CL under the assumption that Wilks' theorem holds (solid lines).
They are obtained by using as ``data'' the no-oscillation prediction without statistical fluctuations (``Asimov data'') and considering contours of $\Delta \chi^2 = 5.99$.
As can be seen from the figure, the best-fit points always lie very close to the expected 
sensitivity limit in this case and, for a sizeable fraction of the pseudo-experiments simulated, they lie 
\emph{to the right} of the sensitivity curve, if naively computed assuming a $\Delta\chi^2$ for 2~DOF (as is usually the case in the literature). 

\begin{figure}[ht!]
\begin{center}
\includegraphics[width=0.47\textwidth]{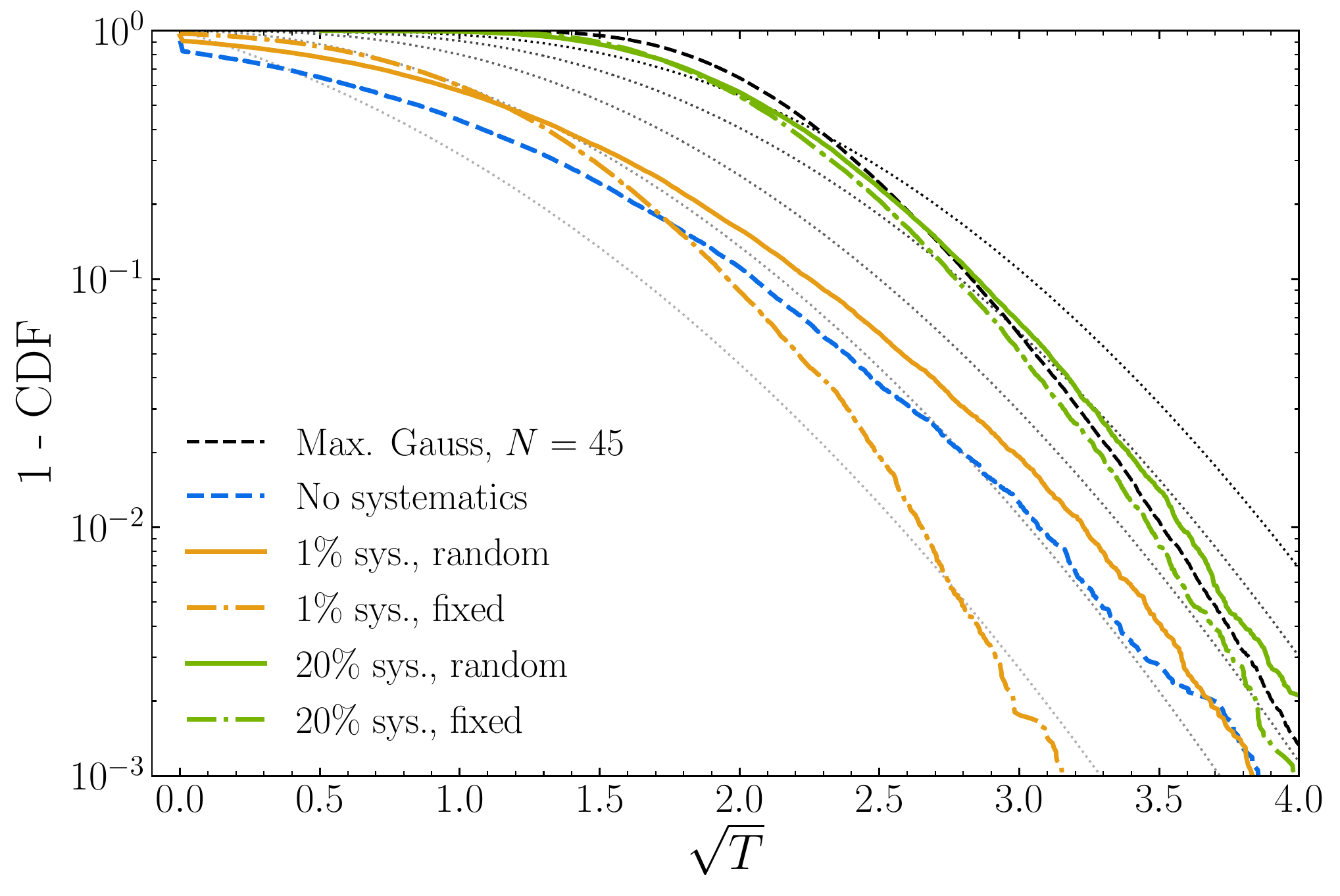}
\caption{\label{fig:random} Impact of a systematic uncertainty on the
  overall normalization. We show the distribution of the square-root
  of the test statistic, $\sqrt{T}$, for no systematic uncertainty,
  1\%, and 20\% uncertainty. The total number of events is
  $\mathcal{N} = 1.5\times 10^4$. For the solid curves we randomize
  the central values for the systematic for each draw of the
  pseudo-data, whereas for the dash-dotted curves the central value is
  kept fixed. The black-dashed curves corresponds to the max.~Gauss
  distribution for $N=45$. The gray dotted lines show the
  $\chi^2$-distributions as the number of degrees of freedom is
  increased from 1 (lightest gray line to the left) to 5 (darkest gray
  line to the right).}
\end{center}
\end{figure}

Let us now discuss the impact of a systematic uncertainty on the overall normalization of the spectrum for the distribution of the test statistic $T$. Fig.~\ref{fig:random} shows the results obtained for different assumed priors on the systematic error for the signal, $\sigma_{\rm sys} = 20\%,~1\%$ as well as the no-systematics case. In all cases we assume a total number of events of $\mathcal{N} = 1.5\times 10^4$. We see from the figure that the distribution for the no-systematic case is somewhat $\chi^2$-like, with a number of DOF between 1 and 2. Although some deviations from this behaviour are observed (due to the effect of the physical boundary $\sin^22\theta \ge 0$ as well as the non-linearity of the model), the distribution is clearly different from a max.~Gauss. The reason is that in Sec.~\ref{sec:toy} we assumed that only shape information is used (assumption~($b$)), whereas in the absence of systematic errors the information on the total event rate is also available. In this case the model of fitting white noise, eq.~\ref{eq:chi2_2}, does not fully correspond to fitting the disappearance probability in eq.~\ref{eq:Posc}, which can only reduce the event numbers. 
In contrast, for the case $\sigma_{\rm sys} = 20$\% the systematic uncertainty is much larger than the statistical one for the assumed event sample, $\sigma_{\rm sys} \gg 1/\sqrt{\mathcal{N}}$. This corresponds effectively to a free normalization in the fit and assumption~($b$) is satisfied. Correspondingly, we observe in Fig.~\ref{fig:random} a very good agreement with the max.~Gauss distribution for this case. \changes{Note that also for the 10\% systematic assumed in Fig.~\ref{fig:bestfit} we have 
$\sigma_{\rm sys} \gg 1/\sqrt{\mathcal{N}}$, such that assumption $(b)$ defined in Sec.~\ref{sec:assumptions} is fulfilled.} For the 1\% case we have $\sigma_{\rm sys} \simeq 1/\sqrt{\mathcal{N}}$, which corresponds to an intermediate situation between fixed and free normalization.

Fig.~\ref{fig:random} also shows the impact due to the treatment of systematics when simulating the random pseudo-data. Solid curves show the results obtained randomizing the central value of the pull parameter, i.e., for each realization of the pseudo-data we draw $\bar\xi$ in eq.~\ref{eq:chi2_total} from a Gaussian distribution with width $\sigma_{\rm sys}$ (as outlined in Sec.~\ref{sec:sim-details}). In contrast, for the dash-dotted curves the central value for the pull parameter is not randomized and kept fixed at $\bar\xi = 0$ for all pseudo-data samples. We see that this has a rather large impact on the $\sqrt{T}$ distribution as long as systematic and statistical \changes{uncertainties} are comparable ($\sigma_{\rm syst} = 1\%$), whereas for an effectively free normalization ($\sigma_{\rm syst} = 20\%$) the difference is largely reduced. The reason is that in the latter case the fit can always adjust the normalization within the statistical uncertainty, with negligible impact of the penalty term for the pull parameter.

\subsection{Further studies of the properties of the {\it T} distribution}

\begin{figure*}[t!]
\begin{center}
\begin{tabular}{cc}
\includegraphics[scale=0.5]{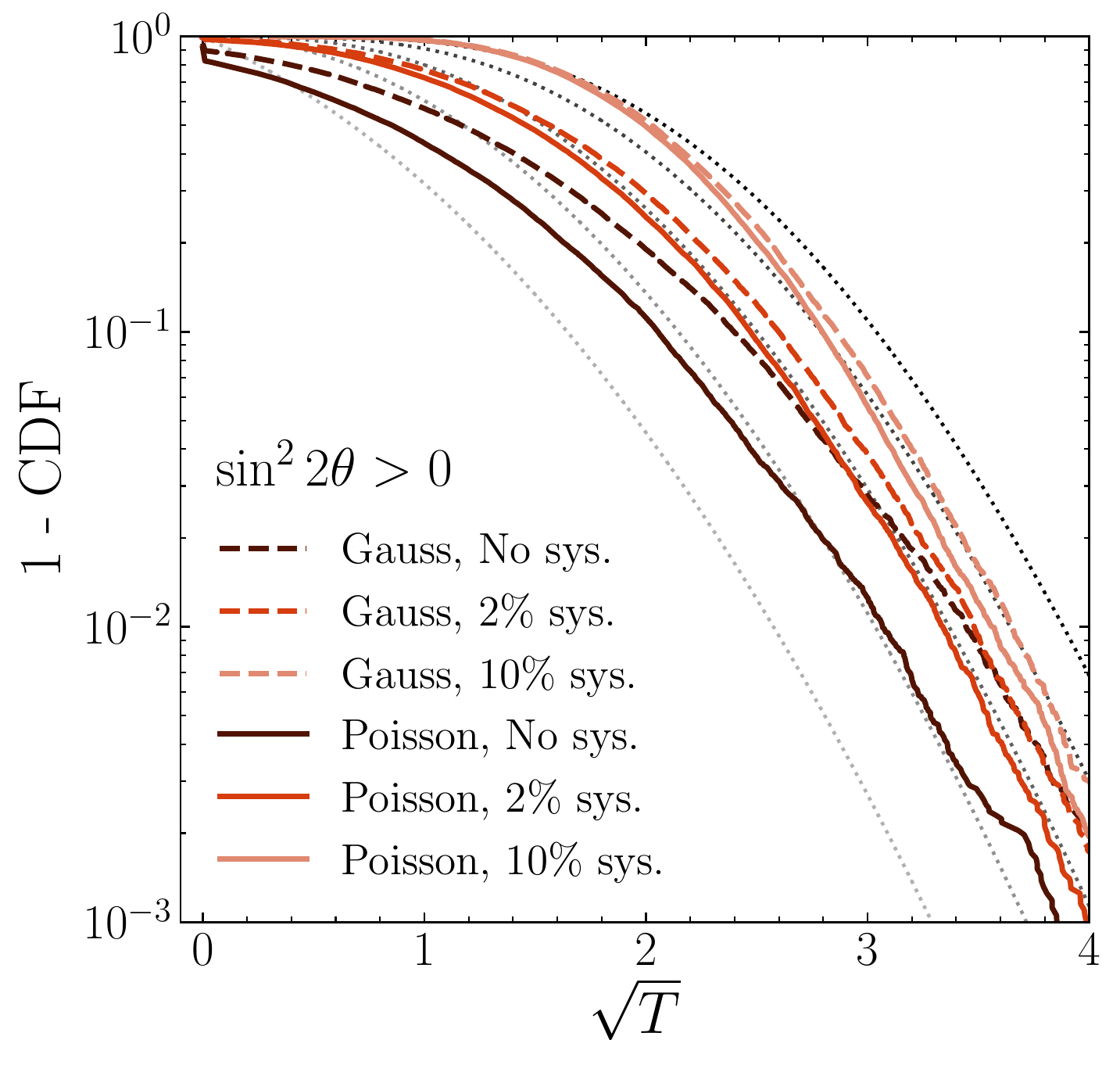} & 
\includegraphics[scale=0.5]{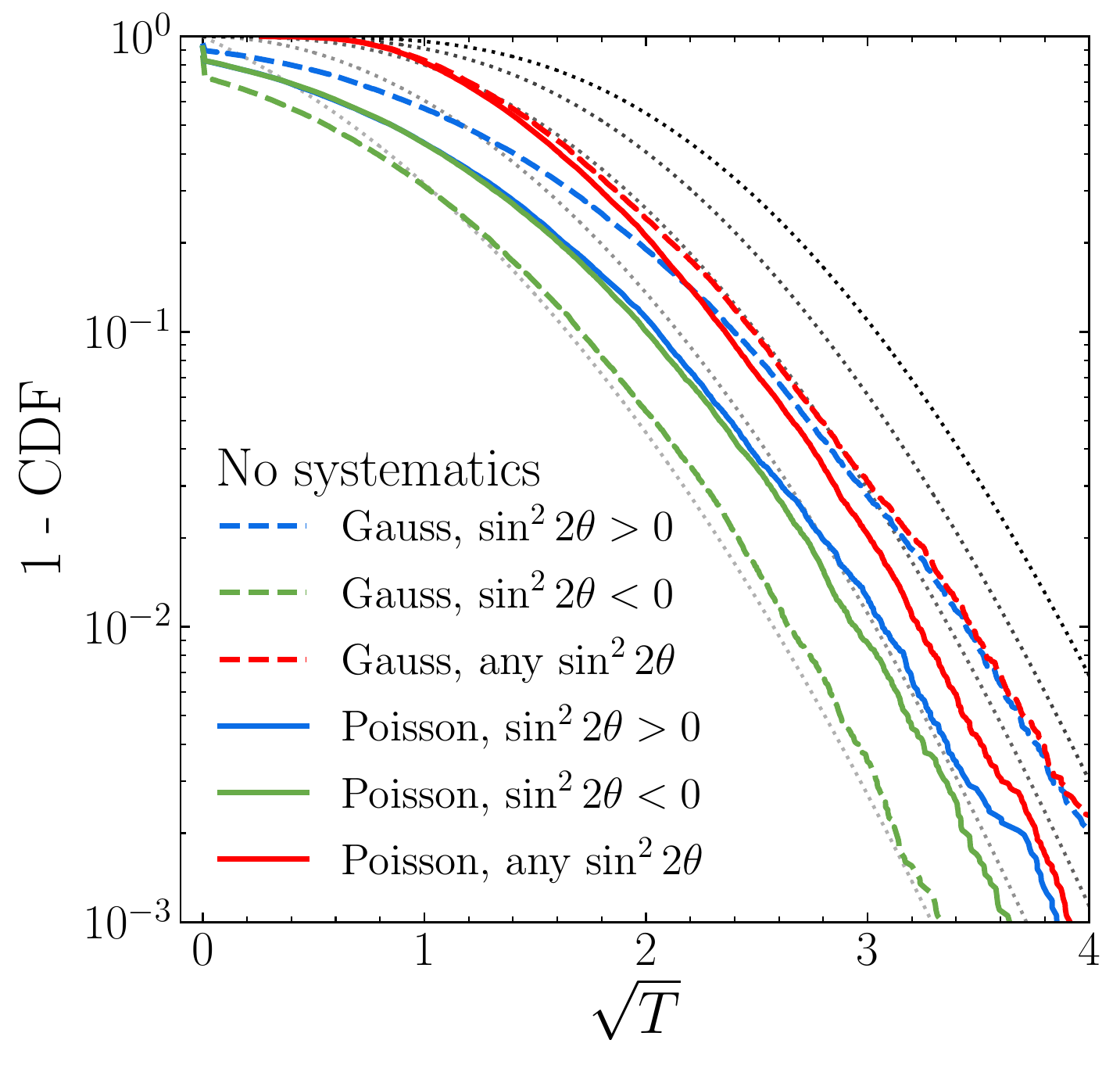} 
\end{tabular}
\caption{\label{fig:poisson-gaus}  
Impact on the distribution of $\sqrt{T}$ due to the choice of the $\chi^2$ implementation. Solid curves correspond to a Poisson $\chi^2$, while dashed lines correspond to a Gauss $\chi^2$. 
The left panel shows the effect of increasing the systematic uncertainties from no systematics to an overall 10\% signal normalization uncertainty.
In the right panel we show the results by imposing different restrictions on the allowed range for $\sin^22\theta$, as indicated by the labels. In all cases, a total of $1.5\times 10^4$ events are simulated for each pseudo-experiment. Dotted curves indicate the $\chi^2$ distribution for 1 to 5 DOF from left to right.}
\end{center}
\end{figure*}

In this subsection we investigate in some detail additional properties of the distribution of the test statistic $T$. We start by discussing the impact of the two $\chi^2$ implementations from eq.~\ref{eq:chi2-stat-poisson} (Poisson $\chi^2$) versus
eq.~\ref{eq:chi2-stat-gauss} (Gauss $\chi^2$). Naively one expects that they should give similar results if the number of events per bin is $\gtrsim 10$. The differences on the resulting distributions for the two $\chi^2$ implementations are shown in Fig.~\ref{fig:poisson-gaus}.

In the left panel of Fig.~\ref{fig:poisson-gaus} we adopt different choices for the normalization uncertainty. Interestingly, we find for $\sigma_{\rm sys} \lesssim 2\%$ notably differences, despite the rather large event number of $\mathcal{N} = 1.5\times 10^4$. For the 43 bins in our simulation this corresponds to about 350 events per bin on average, where the bin with the smallest number of events has a mean above 20 events.
We have checked that for the no-systematic case the differences between Poisson and Gauss disappear only for $\mathcal{N} \gtrsim 10^5$. 
From the figure we also see that for $\mathcal{N} = 1.5\times 10^4$ the differences between the Gauss and Poisson implementations disappear for large enough systematic uncertainty, when both cases approach the max.~Gauss distribution. 
The origin of the different behavior is related to the assumption $\sigma_i = \sqrt{d_i}$ in eq.~\ref{eq:chi2-stat-gauss}. We have confirmed that when we use instead the \changes{Pearson definition, eq.~\ref{eq:chi2-stat-pearson}, with $\sigma_i = \sqrt{p_i^0}$, the Pearson} and Poisson $\chi^2$ implementations lead to identical results. 

In the right panel of Fig.~\ref{fig:poisson-gaus} we study the impact of the physical boundary $\sin^22\theta \ge 0$ in the case of $\mathcal{N} = 1.5\times 10^4$ and no systematic uncertainty on the normalization. \changes{Note that, while only $\sin^22\theta \ge 0$ makes sense from the mathematical point of view, this parameter controls the amplitude of the oscillation and in principle it is possible to try to fit data without taking this requirement into account. Thus, } in the figure we compare three cases:  
$\sin^22\theta \ge 0$ (blue lines), $\sin^22\theta \le 0$ (green lines), and a third case where no restriction is imposed on $\sin^22\theta$ (red lines). For the Poisson $\chi^2$ implementation we see that restricting the sign of $\sin^22\theta$ has a notable impact on the distribution, but the effect is similar regardless of the sign of $\sin^22\theta$. In contrast, for the Gauss $\chi^2$ implementation we observe significant differences between the cases 
$\sin^22\theta \ge 0$ and $\sin^22\theta \le 0$. Again this is a consequence 
of using $\sqrt{d_i}$ as the statistical \changes{uncertainty} in the Gauss $\chi^2$, eq.~\ref{eq:chi2-stat-gauss}: 
as $d_i$ includes statistical fluctuations, $\sqrt{d_i}$ is not symmetric between upward and downward fluctuations, which leads to the asymmetric behavior with respect to the sign of $\sin^22\theta$. 
\changes{In contrast, if the theoretical prediction is used as variance as in the Pearson definition, eq.~\ref{eq:chi2-stat-pearson}, the $\chi^2$} becomes symmetric between upward and downward fluctuations. We have explicitly checked that in this case the dependence on the sign of $\sin^22\theta$ disappears and we recover the result from the Poisson $\chi^2$.
Surprisingly, these second order effects are not negligible even for $\mathcal{N} = 1.5\times 10^4$ events. 

Overall, we observe 
\changes{that the results agree with the Poisson $\chi^2$ if we use the 
the Pearson $\chi^2$, where the
square-root of the prediction to calculate the statistical uncertainty, while sizeable deviations occur for the Gaussian $\chi^2$, where the square-root of the data is used as statistical uncertainty.} Let us remark, however, that as long as the distribution of the test statistic is numerically evaluated by Monte Carlo simulation, of course any reasonable $\chi^2$ definition can be used (including also the Gauss $\chi^2$ as defined in eq.~\ref{eq:chi2-stat-gauss}).

To summarize the results found in this section, we find that as long as $\sigma_{\rm sys} \lesssim 1/\sqrt{\mathcal{N}}$, the distribution of the test statistic $T$ is sensitive to details of the analysis, such as size of systematics, treatment of systematics during randomization, $\chi^2$ variants, physical boundaries. However, once $\sigma_{\rm sys} \gg 1/\sqrt{\mathcal{N}}$ (i.e., for experiments where only shape information is used) the max.~Gauss distribution seems to be a rather robust result.

\section{Application: Neutrino-4 as a case study}
\label{sec:neutrino4}

The Neutrino-4 experiment \cite{Serebrov:2020kmd,Serebrov:2018vdw} has recently claimed a possible indication of sterile neutrino oscillations with $\Delta m^2 \simeq 7.2$~eV$^2$ and $\sin^22\theta \simeq 0.26$. They report a statistical significance using their combined phase 1 and 2 data of 3.2$\sigma$~\cite{Serebrov:2020kmd}. Then an estimate of their systematic uncertainty is quoted, leading to a combined statistical/systematical significance of a positive $\sin^2 2\theta$ of $2.8\sigma$. In this section we use the Neutrino-4 results to illustrate the arguments presented above on a real-life example.
We concentrate on the purely statistical aspect and will show that, in light of the discussions in the previous sections, the significance from statistical \changes{uncertainties} alone is already lower than the quoted 3.2$\sigma$.

Neutrino-4 uses a segmented detector, which allows to bin their data in both $L$ and $E$. The data is binned using 9 bins in energy with width $\Delta E = 0.5$~MeV starting at 2.3~MeV, and 24 bins in baseline with $\Delta L = 0.235$~m starting at 6.25~m, resulting into a total of 216 bins in $L/E$. 
The bin width in energy of 0.5~MeV corresponds to the energy resolution of the detector~\cite{Serebrov:2020kmd}. 
Eventually, each consecutive group of 8~bins are combined together, leading to $N=27$ data points. The observed data correspond to the ratio
\begin{equation}\label{eq:Robs}
    R_i^{\rm obs} = \frac{d_i}{\frac{1}{N}\sum_{i=1}^N d_i} \,,
\end{equation}
where $d_i$ stands for the observed number of events in bin $i$. Out of these, the first 19 bins are shown for the combined phase~1 and phase~2 data sets in Fig.~47 of \cite{Serebrov:2020kmd} (blue points). Following the Neutrino-4 collaboration, we fit these 19 data points with the survival probability for a given $L/E$ bin over the averaged probability:
\begin{equation}\label{eq:Rpred}
    R_i^{\rm pred} = \frac{1-\sin^22\theta \left\langle\sin^2\frac{\Delta m^2L}{4E}\right\rangle_i} {1-\frac{1}{2}\sin^22\theta} \,.
\end{equation}
Here $\langle\,\cdot\,\rangle_i$ indicates the average over an energy interval $\Delta E$, with 
the value of $L/E$ set at the bin center of the corresponding $(L/E)$-bin $i$. 
The fit is performed with a simple Gaussian $\chi^2$ definition, using the statistical \changes{uncertainties} read off from Fig.~47 of \cite{Serebrov:2020kmd}. Note that, due to the particular way the fit is performed by Neutrino-4, using the ratios in eqs.~\ref{eq:Robs}, \ref{eq:Rpred} the analysis is only sensitive to spectral distortions in $L/E$, and therefore assumption~($b$) from Sec.~\ref{sec:assumptions} is fulfilled.

With our fit we can reproduce to good accuracy the results from Ref.~\cite{Serebrov:2020kmd}. Our best-fit point is located at $\Delta m^2 = 8.84~\mathrm{eV}^2$, $\sin^22\theta = 0.42$; however, we find a quasi-degenerate local minimum 
with $\Delta\chi^2 = 5\times 10^{-3}$ at $\Delta m^2 = 7.28~\mathrm{eV}^2$, $\sin^22\theta = 0.34$, close to the best-fit point obtained by Neutrino-4. We explain this slight difference by the fact that the fit reported in Ref.~\cite{Serebrov:2020kmd} uses more information in $L/E$ than available to us. This additional information seems to somewhat disfavor the local minimum around $\Delta m^2 \simeq 9$~eV$^2$ compared to the one at $\simeq 7.25$~eV$^2$. Furthermore, we obtain for the $\chi^2$ minimum and the test statistic $T$, i.e., the $\Delta\chi^2$ between no oscillations and the best-fit point:
\begin{align}
    &\chi^2_{\rm min} = 16.05 \,,\quad T = 12.94 \quad \text{(our result)} \,, \\
    &\chi^2_{\rm min} = 17.11 \,,\quad T = 12.87 \quad \text{(Fig.~47 of Ref.~\cite{Serebrov:2020kmd})} \,,
\end{align}
showing good agreement, especially for $T$. 

\begin{figure}[t!]
\begin{center}
\includegraphics[width=0.47\textwidth]{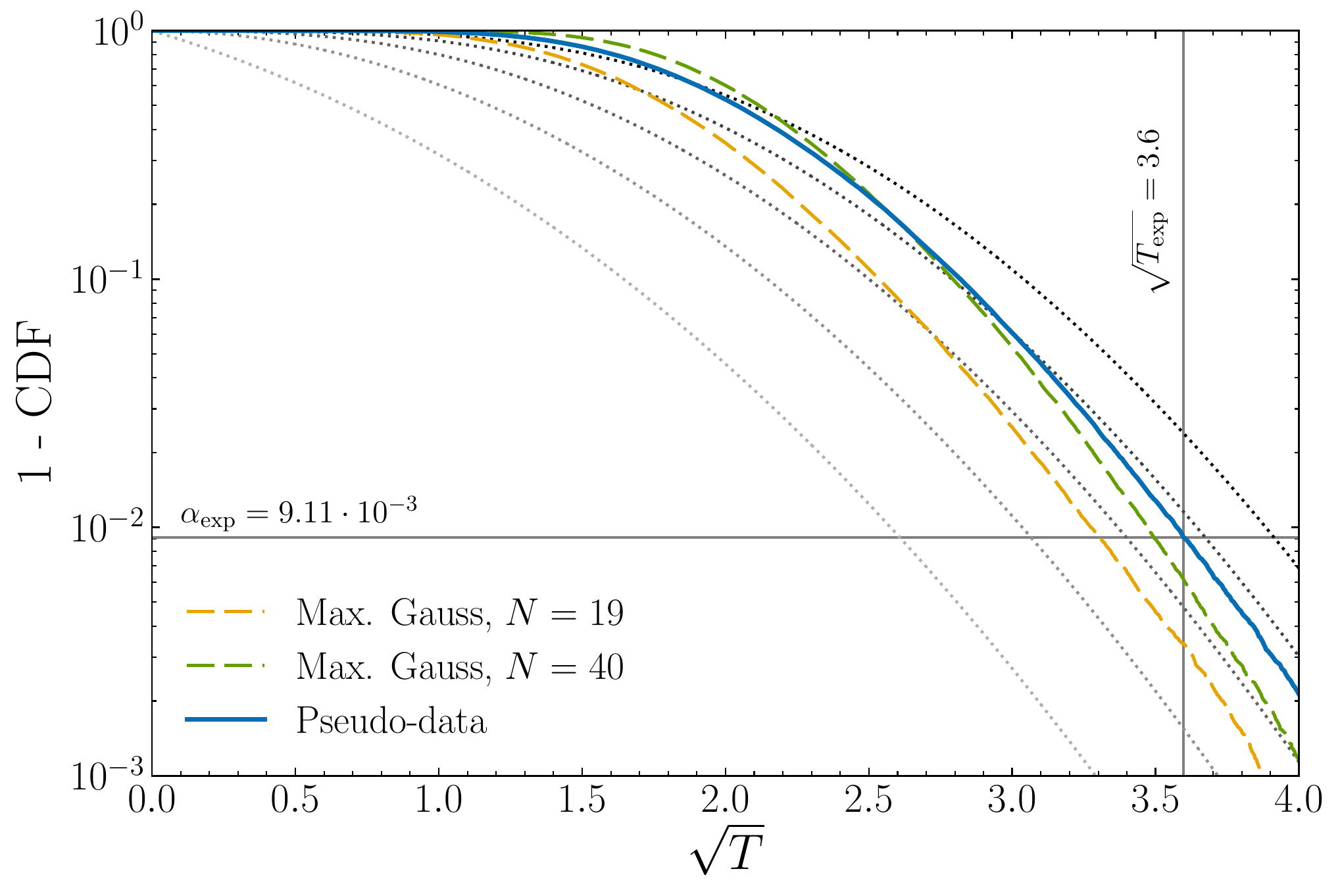}
\caption{\label{fig:neutrino4-T}  Distribution of the the square-root of the test statistics $\sqrt{T}$ for the Neutrino-4 experiment, obtained from the simulation of 100,000 pseudo-data sets under the assumption of no oscillations. The numerical result (solid blue curve) is compared to the expectation for a $\chi^2$ distribution with $1\ldots 5$~DOF (dotted curves, from left to right), as well as to the distribution for the max.~Gauss distribution with 19 and 40 bins (dashed curves). The vertical line indicates the value of $\sqrt{T}=3.6$ obtained from the observed Neutrino-4 data, whereas the horizontal line shows the corresponding $p$-value.}
\end{center}
\end{figure}

If evaluated under the assumption of Wilks' theorem with a $\chi^2$ distribution with 2~DOF we would get from $T = 12.9$ a $p$-value of $1.58\times 10^{-3}$, corresponding to $3.16\sigma$. In order to check this reasoning, we have 
generated a large sample of artificial data sets for Neutrino-4, under the null-hypothesis of no oscillations, in order to calculate the distribution of $T$ explicitly. The result is shown in Fig.~\ref{fig:neutrino4-T}, which shows significant deviations from a $\chi^2$-distribution. In agreement with the discussions above, the distribution of $\sqrt{T}$ is found to be more similar to the max.~Gauss distribution. In this case we find that the effective $N=40$ for the max.~Gauss distribution providing the closest fit to the numerical distribution deviates substantially from the actual number of bins in the data, $N=19$. Based on the numerical $T$ distribution we obtain a $p$-value of $9.1\times 10^{-3}$ (or $2.6\sigma$), indicating that the actual statistical significance is clearly lower. \changes{Note that here we study the significance based on statistical uncertainties only. Systematic effects as mentioned in Ref.~\cite{Serebrov:2020kmd} will reduce the significance even further. See also the discussion in Ref.~\cite{Almazan:2020drb} in this context.}

In addition, we have calculated confidence regions in the plane of $\sin^22\theta$ and $\Delta m^2$ 
by performing a Feldman-Cousins analysis \cite{Feldman:1997qc}. 
For given values of $\sin^22\theta$ and $\Delta m^2$ we have generated many artificial data sets, assuming that these values are the true values in Nature. This allows us to compute the correct distribution of 
\begin{equation}
    \Delta\chi^2(\sin^22\theta,\Delta m^2) = \chi^2(\sin^22\theta,\Delta m^2) - \chi^2_{\rm min} \,
\end{equation}
for each point in the parameter space. Comparing the value of $\Delta \chi^2_{\rm exp}$ obtained from the actual experimental data to the numerical distribution for $\Delta\chi^2(\sin^22\theta,\Delta m^2)$, we obtain the confidence level (CL) at which a particular point can be rejected. Repeating this procedure for the whole parameter space we obtain confidence regions at a given CL as the set of all points which are accepted at that CL. 

\begin{figure}[t!]
\begin{center}
\includegraphics[width=0.47\textwidth]{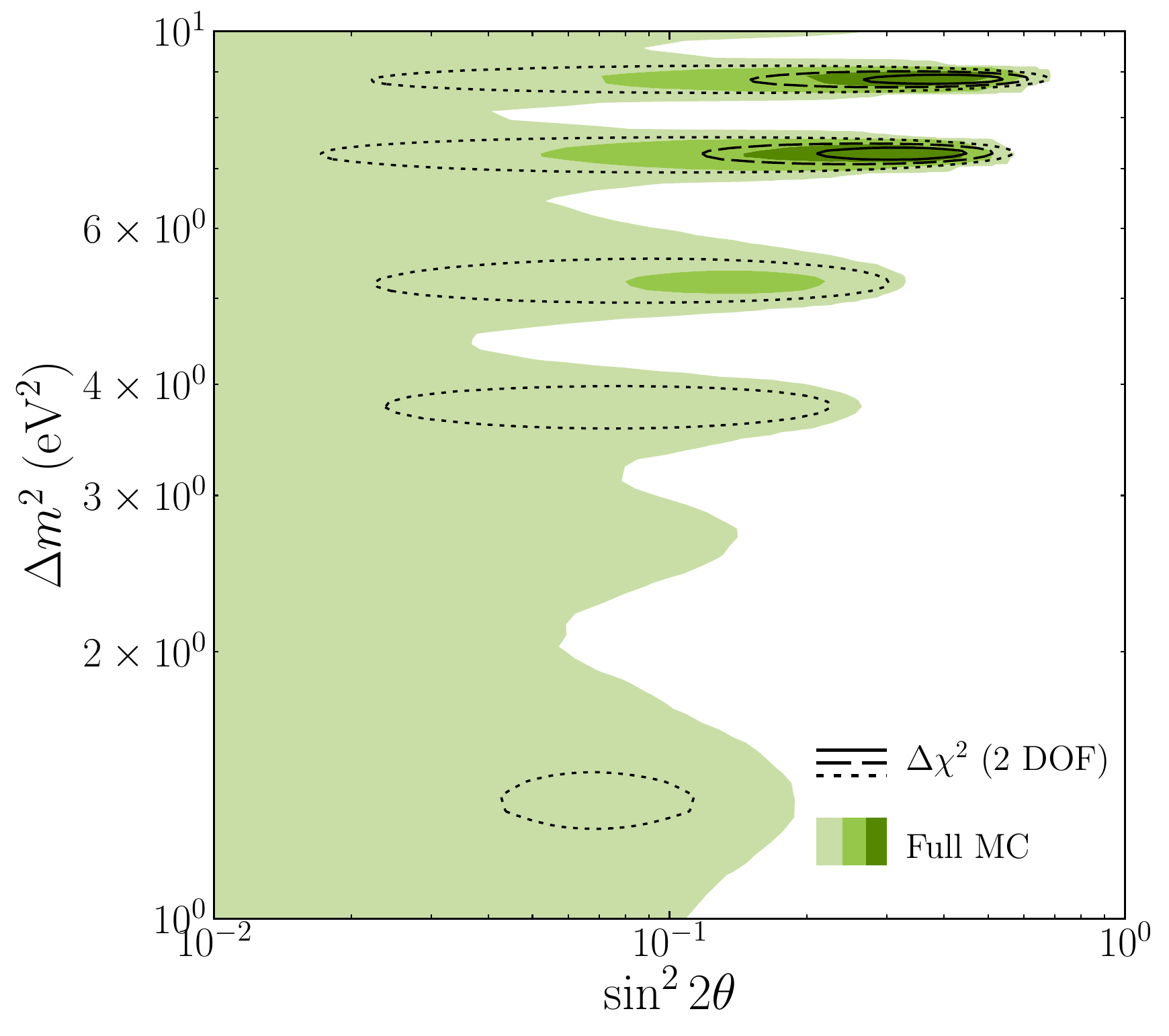}
\caption{\label{fig:neutrino4-cont} Confidence regions from our re-analysis of Neutrino-4 data \cite{Serebrov:2020kmd} at 68.3\% (dark green), 95.45\% (medium green), and 99.73\%~CL (light green). Shaded regions correspond to the confidence regions constructed by Monte-Carlo simulations following the Feldman-Cousins prescription~\cite{Feldman:1997qc}, whereas black curves show the corresponding CL contours in $\Delta\chi^2$ assuming it follows a $\chi^2$-distribution for 2~DOF, that is, $\Delta\chi^2 = 2.3$ (solid), 6.18 (dashed), 11.83 (dotted).}
\end{center}
\end{figure}

The results of this analysis are shown in Fig.~\ref{fig:neutrino4-cont} as shaded regions for 68.3\% (dark green), 95.45\% (medium green), and 99.73\% (light green)~CL. Our regions are also compared to $\Delta\chi^2$ contours obtained under the assumption that $\Delta\chi^2$ follows a $\chi^2$ distribution with 2~DOF, which would be the case if Wilks' theorem held. We clearly observe that true confidence regions are substantially larger than the ones based on the $\chi^2$ distribution. In particular, the no-oscillation case is contained in the $3\sigma$ contour for the Monte-Carlo calculation, in agreement with the discussion of the test statistic $T$ above.\footnote{Let us note that our $\Delta\chi^2$ contours are also somewhat larger than the ones shown in Fig.~45 of Ref.~\cite{Serebrov:2020kmd}. We believe that the reason for this difference is that contours in Ref.~\cite{Serebrov:2020kmd} are drawn for a $\chi^2$ distribution with 1~DOF, while ours are shown for 2~DOF. We have checked that using the same prescription we can reproduce their regions with good accuracy.}

\section{Summary and conclusions}
\label{sec:conclusions}

In this paper we have studied the statistical interpretation of sterile neutrino 
oscillation searches in the disappearance mode, specifically when no
information on the absolute normalization of the signal is used. \emph{A priori} there
are several good reasons to expect that Wilks' theorem does not apply in this case:
the presence of a physical boundary, the fact that the parameter space changes dimension
if either $\Delta m^2 \to 0$ or $\sin^22\theta \to 0$, and the highly non-linear 
dependence of the number of events on $\Delta m^2$. Not surprisingly, we
do indeed find significant deviations. Although in this work we decided to focus on 
short-baseline reactor experiments as a case study, our results are more general. We find that
this situation, under some assumptions, is equivalent to fitting
Gaussian white noise with a single frequency of free amplitude. This allows us to express the distribution of the test statistic
$T$ to be the maximum of $N$ Gaussian random variables where $N$ is
the effective number of bins (``max.~Gauss distribution''). Therefore, this class of oscillation
searches will always find a best-fit for a non-zero signal even if there is no
oscillation in the data, with a non-negligible
statistical significance if interpreted as if Wilks' theorem would apply. In other words, the parameters obtained at the minimum of the $\chi^2$ are biased estimators in this case. 

We then perform Monte Carlo simulations of a toy reactor disappearance experiment, to confirm that
our analytic understanding carries over to a more realistic setting. 
The test statistic $T$ we consider is equivalent to the log-likelihood either
for a Gaussian likelihood or Poissonian likelihood. We find that, if the systematic uncertainty on the event normalization is comparable to (or smaller than) the statistical \changes{uncertainty} of the event sample, 
the distribution function of $T$ is rather sensitive on fine details of the chosen simulation and, in particular, on: whether 
the central value of the nuisance parameter is randomized or not, and
whether a Gaussian or a Poissonian log-likelihood is used (despite the
fairly large number of events per bin).
Conversely, for experiments relying on shape information only (that is, when no information on the absolute normalization is used) the max.~Gauss distribution is a rather robust prediction for the distribution of the test statistic. \changes{Although the shape of the max.~Gauss does not depend strongly on the value of $N$ (only logarithmically), we have not found a simple way to predict the value that provides the best description of the distribution of the test statistic.  }

Finally, we apply our understanding to the actual data of the
Neutrino-4 experiment. We are able to reproduce the quantitative
details of their analysis quite well if we assume that Wilks' theorem
applies. However, in agreement with our arguments presented above we find 
 that the test statistics shows significant deviations from a $\chi^2$ distribution. 
In particular, we show by
explicit Monte Carlo simulation that the significance
of the claimed oscillation signal is reduced from 3.2\,$\sigma$
($p=1.58\times 10^{-3}$) to $2.6\,\sigma$ ($p=9.1\times10^{-3}$), that
is, the probability that this is a mere statistical fluctuation is
about 6 times larger than that expected if Wilks' theorem were to hold.
It should be noted that our Neutrino-4 analysis is based on statistical \changes{uncertainties} only, and that the inclusion of systematic effects may reduce the significance even further.

In summary, our results provide a simple, intuitive understanding on
why and how shape-only oscillation searches are different from the
usual case. Applied to Neutrino-4 we find a reduced significance for
sterile neutrino oscillation, but not to the extent to completely
dismiss this indication as a pure statistical fluctuation. It would be
interesting to see how this type of analysis would play out in a
global fit of all short-baseline reactor data, but this is beyond the
scope of the present work.

\begin{acknowledgement}
The authors warmly thank Mattias Blennow and Enrique Fer\-nandez-Martinez for useful discussions. PH acknowledges support from the U.S. Department of Energy Office of Science under contract \protect{DE-SC0018327}. PC acknowledges support from the grant \protect{PROMETEO/2019/083}, from the Spanish MICINN through the ``Ram\'on y Cajal'' program under grant \protect{RYC2018-024240-I}, and from the Spanish Agencia Estatal de Investigacion through grant ``IFT Centro de Excelencia Severo Ochoa SEV-2016-0597''. The authors also acknowledge use of the HPC facilities at the IFT (Hydra cluster) and IFIC (SOM cluster). This work was partially supported by the European projects H2020-MSCA-ITN-2015//674896-ELUSIVES and 690575-In\-visiblesPlus-H2020-MSCA-RISE-2015.
\end{acknowledgement}

\bibliographystyle{JHEP_improved}
\bibliography{refs}

\end{document}